\pdfoutput=1

\documentclass[11pt]{article}

\usepackage{EMNLP2022}

\usepackage{times}
\usepackage{latexsym}
\usepackage{amsmath}
\usepackage{amsfonts}

\usepackage[T1]{fontenc}

\usepackage[utf8]{inputenc}

\usepackage{microtype}

\usepackage{inconsolata}

\usepackage{graphicx}

\title{Dense Feature Memory Augmented Transformers for COVID-19 Vaccination Search Classification}

\author {
Jai Gupta\\
Google Research\\
jaigupta@google.com

\And Yi Tay\\
Google Research\\
yitay@google.com

\And Chaitanya Kamath\\
Google Research \\
ckamath@google.com

\And Vinh Q. Tran\\
Google Research \\
vqtran@google.com
\AND
Donald Metzler\\
Google Research \\
metzler@google.com
\And Shailesh Bavadekar\\
Google Research \\
shaileshb@google.com
\And Mimi Sun\\
Google Research\\
mimisun@google.com
\And Evgeniy Gabrilovich\\
Google Research\\
gabr@acm.org
}

\begin{document}
\maketitle
\begin{abstract}
With the devastating outbreak of COVID-19, vaccines are one of the crucial lines of defense against mass infection in this global pandemic. Given the protection they provide, vaccines are becoming mandatory in certain social and professional settings. This paper presents a classification model for detecting COVID-19 vaccination related search queries, a machine learning model that is used to generate search insights for COVID-19 vaccinations. The proposed method combines and leverages advancements from modern state-of-the-art (SOTA) natural language understanding (NLU) techniques such as pretrained Transformers with traditional dense features. We propose a novel approach of considering dense features as memory tokens that the model can attend to. We show that this new modeling approach enables a significant improvement to the Vaccine Search Insights (VSI) task, improving a strong well-established gradient-boosting baseline by relative +15\% improvement in F1 score and +14\% in precision. 
\end{abstract}

\section{Introduction}

\begin{figure*}[t]
  \includegraphics[height=5cm]{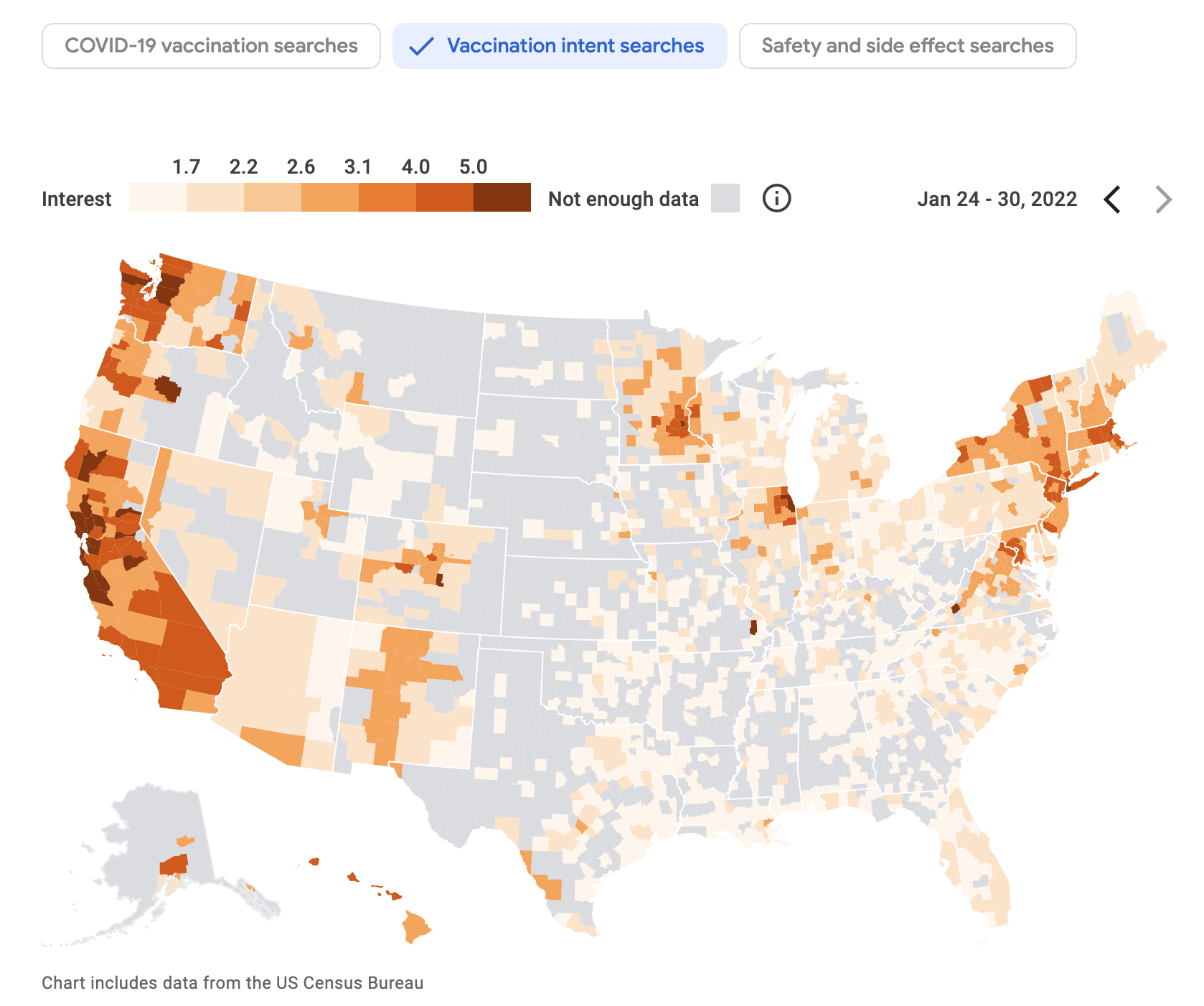}
  \includegraphics[height=5cm]{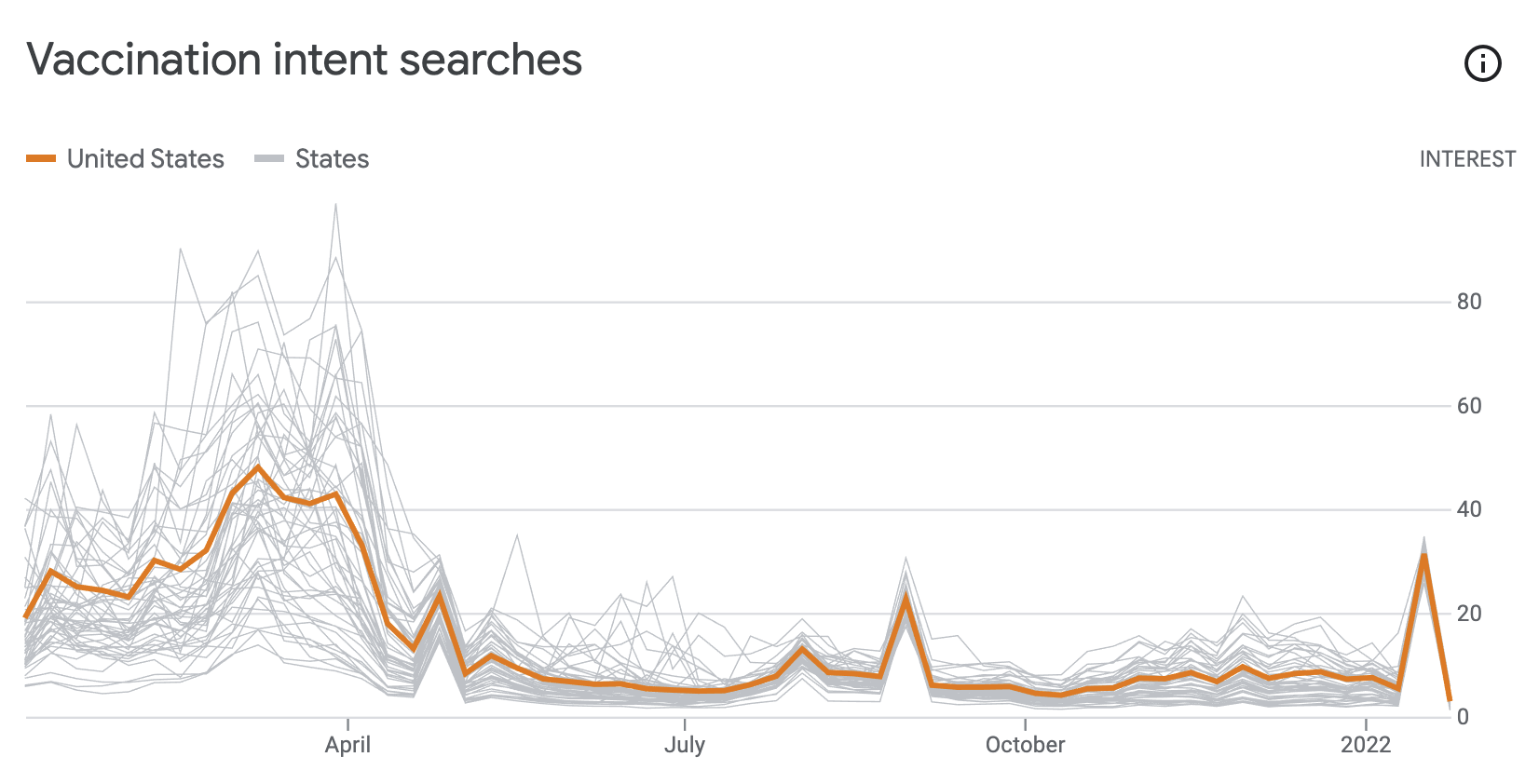}
  \caption{VSI tool presenting vaccination intent search query statistics segmented by location and time respectively.}
    \label{fig:vsitool}
\end{figure*}

Though COVID-19 continues to be a challenge worldwide, vaccines have provided the much needed hope. Countries and governments have significantly ramped up their efforts to improve the reach of the vaccines including booster shots. As such, it is important to understand how users search for vaccine related information such as vaccine efficacy, safety, and regional availability as this information can be very useful to inform policy decision making, create effective public service announcements, implement more efficient distribution of vaccines, etc. To this end, we released a public tool for COVID-19 Vaccination Search Insights, an interactive report that provides insights on user searches for COVID-19 vaccinations. For an example please see Figure \ref{fig:vsitool}. On the backend, this tool performs privacy preserving classification of user search queries and clusters them based on the region they were issued from to present a timeline of how these searches have changed over time. A core task in the VSI tool is the problem of classifying search query intent, and one of them is whether users are seeking information on \textit{vaccine access}, i.e. queries related to the eligibility, availability, and accessibility of COVID-19 vaccines.

The task is a challenging one since simply matching for COVID related terms is insufficient, as queries such as \textit{fully vaccinated travel} or \textit{proof of covid vaccination} are negative classes. Hence, this problem is nuanced and may benefit from a coalition of advanced language understanding systems and traditional search-related feature engineering methods. As such, the problem at hand crosses two main modalities, i.e., text and traditional dense features. In this problem, the text features are user search queries that are short and/or lack context. The dense features (discussed in more detail in section \ref{sec:dataset}) are hand crafted features from recent and past activities, previous clicks, and named entities that play a critical role in understanding the user's intent. However, dense features alone fail to capture important contextual language cues, such as those that state-of-the-art natural language understanding systems~\citep{bert,t5} have been shown to handle well.
We find that both modalities are highly complementary and it is difficult to achieve strong performance using only a single modality. 



\paragraph{Our Contributions} The overall contributions of this paper can be summarized as follows:
\begin{itemize}
    \item We propose a new model and framework for search query intent classification for our COVID-19 Vaccination Search Insight tool.
    \item We propose a paradigm of exploiting the benefits of text inputs through state-of-the-art NLU models, along with traditional dense features found in large-scale systems. 
    \item We propose a novel method of fusing dense features with Transformers that enables queries to retrieve from a dense memory store, in similar spirit to a contextual key-value store.
    Memory tokens here are used in similar manner as memory-based methods described in \citep{tay2020efficient} and models such as Set Transformer \citep{lee2019set}, Memory Transformers \citep{sukhbaatar2019augmenting, memorytransformer, memformer}, and Global Memory tokens in ETC \cite{etc} and BigBird \cite{zaheer2020big}. Notably, this is the first proposal for constructing global memory tokens using dense features.
    \item We conduct extensive experiments on real production data from three geographical regions. Our experiments show that the proposed method significantly outperforms a strong gradient boosting baseline by $+15\%$ and outperforms a SOTA Transformer by $+5\%$ on F1 score achieving very strong F1 score of over $98\%$ on the US dataset with similar strong performance for other regions. 
\end{itemize}

\section{Related Works}
This section presents related works and the background for this paper.

\textit{Classification with Feature-based ML.} Building machine learned (ML) systems that operate across dense hand-crafted features is a well-established method. It is common to consider this class of ML methods as \textit{tabular} machine learning whereby tree-based methods \citep{chen2016xgboost,adaboost} are dominant\footnote{\url{https://www.kaggle.com/shivamb/data-science-trends-on-kaggle}}. Within the context of text classifiers, feature engineering typically leverages  stemming, lemmatization, part-of-speech tags, tf-idf vectors, entities, salient terms, and other features that are relevant to the task at hand. It is also popular to use semantic representations from Glove \citep{pennington2014glove} or BERT \citep{bert} as input dense features to a model.

\textit{NLU with Pretrained Transformers.} Transformers \citep{vaswani}, characterized by interleaved self-attention and MLP blocks, have become the dominant sequence model for language processing and understanding (NLU) \citep{bert,t5,brown2020language}. The key idea behind self-attention is to perform token-to-token alignment where the joint interaction of queries and keys retrieve from a memory store (value). To this end, it is also common for advanced Transformer architectures to leverage global memory tokens \citep{zaheer2020big,lee2019set,jaegle2021perceiver} that act as a parameter store for the query to attend to \citep{tay2020efficient}.
A cornerstone of these systems is the pretraining task that learns general purpose language representations, which have been shown to be extremely beneficial \citep{tay2021pre} due to the gains from transfer learning \cite{transferlearn}. 

\textit{Joint Learning of Textual and Dense Features.} Performing feature extraction to convert text into dense features (e.g. TF-IDF \cite{tfidf}, word2vec \cite{word2vec}, etc.) for the purpose of learning classifiers jointly with other non-textual (numerical) dense features has been common practice in machine learning for some time \cite{textsurvey, emailvalet, richardson-ads}.
 Previous research in the multi-modal domain adopt a strategy of \textit{early fusion} or \textit{late fusion}: joining the two modalities (in this case textual and numerical) either in feature space early in the architecture or in semantic/decision space late in the architecture, respectively \cite{fusion}.
 Perhaps most related to our approach is a recent work on joint representation of text and tabular data \cite{tabert, tat-qa} that pass a flattened representation of a table alongside text during encoding. Our work instead provides a method for jointly encoding text with dense features of a more generic, unstructured form.
 

\textit{COVID-19 Vaccine Search Insights.} An important distinction of Query classification from generic text classification is that the former are significantly shorter, and may be underspecified \cite{detectingsession, Beitzel2005AutomaticWQ}. Due to this nature, some previous works in this area have augmented queries with additional context to improve query classification performance \cite{broder-rare, shen-bridge, li-query}.

Analyzing user queries and social interactions in health settings has been well studied. \citet{lymelight} presents an analysis of user search queries for Lyme disease forecasting. Similarly, \citet{finder} uses a machine-learned model for real-time detection of foodborne illness using web search and location data. 



\section{Problem Description}
Given a search query $q$, the objective is to classify whether the query was issued with the intent of seeking information related to \textit{vaccine access}.
Notably, $q$ is short in length and may or may not contain all the information needed to make the correct prediction.

Additionally, each $q$ is supplemented with numerical features in the form of a dense feature vector $X_f \in \mathbb{R}^{d_{features}}$, where $d_{features}$ can be any number of features. Details about how $X_f$ is constructed for our setup is present in section \ref{sec:dataset}, but in short  $X_f$ represents the topicality scores of phrases related to the query with $d_{features}$ being 60k. However, it is important to note that all methods described in this paper are agnostic to the source of $X_f$ and can be applied to any vector of numerical features.


\section{VSI Transformer}
This section describes the proposed method. 

\subsection{Pretrained Transformer Encoder}
The main backbone of the proposed architecture is a Transformer~\citep{vaswani} encoder. We leverage the state-of-the-art T5 \citep{t5} model as a starting point.
Since T5 is a seq2seq model, we only utilize the T5 encoder as the Transformer model and discard the decoder of the pretrained model for our classification task. This is done by pooling the output of the encoder stack followed by a dense classification layer.
\subsection{Input formulation}
Given a query $q$, the input to the model is a discrete integer sequence representing the tokens \cite{sentencepiece} of $q$, i.e., $X_{\ell}$ where $\ell$ is the number of tokens in the query from the subword vocabulary $V$. The input sequence is selected from an embedding matrix of $\mathbb{R}^{|V| \times d_{model}}$ to form a tensor of $\mathbb{R}^{\ell \times d_{model}}$.
\subsection{Dense Feature Memory}
For each input-target example, the input to the Dense Feature Memory module is a dense feature $X_f \in \mathbb{R}^{d_{features}}$. Given this dense feature of dimensions $d_{features}$, we transform it into memory tokens of dimension $d_{model}$ via:
\begin{align*}
M_{i} = ReLU(W_iX_{f} + b_i)
\end{align*} where $W_{i} \in \mathbb{R}^{d_{features} \times d_{model}}$ and $M_{i}$ is the i-th memory token. We consider the number of memory tokens to be a hyperparameter. To this end, we then concatenate $[M_1; \cdots M_{N_{memory}}]$ to the input query sequence $X \in \mathbb{R}^{\ell \times d_{model}}$. Along with the input query, we also pass in dense features corresponding to the query to the main body of the network. We note that memory tokens participate in the rest of the computation in a similar spirit to query tokens, i.e., they go through the same MLP and self-attention layers. The dense features are of $d_{features}$ dimensions and are passed into the dense memory module. The dense features used in our setup is explained in section \ref{sec:dataset}.


\subsection{Attention Blocks: Querying \& Retrieving from Dense Feature Memory Tokens}

The dense feature memory token is appended to the input sequence and participates in the self-attention mechanism of the Transformer model. Concretely, the QK matrix of the Transformer can be now written as:
$$A_{\ell,h} = Softmax([Q_{\ell,h};m_{\ell,h}][K_{\ell,h};m_{\ell,h}]^\top_{\ell,h})$$
$$Y_{\ell,h} = A_{\ell,h}[V_{\ell,h};v_{\ell,h}m_{\ell,h}]$$
where $Y_{\ell,h}$ is the $h$-th head of the output at layer $\ell$, $Q,K,V$ are the standard transformations of the query input sequence, and $m_{\ell,h}$ is the dense feature memory token for layer $\ell$. Since $Q$ and $V$ are both augmented with dense features, this provides an opportunity for both the dense features to align with query tokens and vice versa. 

\subsection{Output layer and Optimization}
The final output layer of the Transformer stack is then passed into a pooling and MLP layer.
\begin{align}
Y_{out} = MLP(\psi(Y_{L}))     
\end{align}
where $\psi(.)$ is a pooling operator that maps $\mathbb{R}^{n \times d_{model}} \rightarrow \mathbb{R}^{d_{model}}$. Our $MLP(.)$ function maps $\mathbb{R}^{d_{model}} \rightarrow \mathbb{R}^{N_{class}}$ to the number of classes. Our model then optimizes the Softmax cross entropy loss between the true classes and the predicted values. 
$L = \sum^{L} \sum^n_{i=1} y_i \log(\pi_i) + (1 - y_i)\log(1-\pi_i)$,
where $\pi_{i}$ is the prediction of class $i$ and $y_{i}$ is the ground truth label of the class $i$.

\section{COVID-19 Vaccine Access Dataset}
\label{sec:dataset}
Due to the novel nature of COVID-19, no previous datasets exist to accurately learn a model for the purpose of vaccine access query classification. Thus, in this section we outline the process we used to create this dataset.

\textit{Collection.} To collect a dataset of queries to be labeled for vaccine access, we sample anonymized queries from real search traffic. Since a small minority of our search queries are for COVID-19 vaccination topics,
we leveraged Google's Knowledge Graph entities to find queries that included high confidence positives, potential positives, and close negatives. For example, for high precision candidates we sample top and random queries associated with the entity “COVID-19 Vaccination”, while for high recall low precision candidates, we sample queries that are only associated with the entity “COVID-19” or with “Vaccination”. 

\textit{Labeling.} To label this dataset for the specific purpose of vaccine access, we rely on a large pool of search quality raters who have deep experience with how health-related information needs are reflected in search queries.
These raters were unknown to and independent of the developers of the classifiers. Each query is rated by three independent raters.

\textit{Label Expansion.} We expand our dataset using label propagation to queries that are very similar to labeled queries.
We include examples of positive and negative vaccine access queries in Table \ref{tab:examples}.

\textit{Dense Feature Augmentation.} We augment our dataset by supplementing each query using dense features. To generate these dense features we use a combination of (1) entities mentioned in the query via a proprietary library analogous to Google Cloud Entity Analysis \cite{googlecloudentityanalysis} and (2) related search queries determined by a proprietary algorithm. We pool these two sources and use the 60,000 most common words and phrases to create a dense feature representation with dimension 60,000. At each dimension, we assign a relevance score for the phrase. For mentioned entities, this is analogous to salience in the Google Cloud Entity Analysis API. Table \ref{tab:top-features} shows some of the top features that are generated for each classification split.

    
\section{Experiments}
Below, we dive into experiment setups and results.
\subsection{Baselines}\label{sec:baselines}
We compare our proposed approach with three competitive baselines. The choice of baselines serves two primary purposes, i.e., (1) to show our method is competitive against well-established methods, and (2) to confirm certain scientific hypothesis by ablation-like studies. Please see Section \ref{sec:impldetails} in the Appendix for further implementation details on how we configure and train our models. 


\textbf{Adaboost} A technique used to create an ensemble of weak learners that begins by fitting an estimator on the original dataset and then repeatedly fits additional estimators focusing more on examples that are misclassified by the combination of all the existing set of estimators.
Mathematically, the ensembled AdaBoost classifer can be represented as: $F_N(x) = \sum_{n=1}^N f_n(x)$ which consists of $N$ weak learners ($f_n(x)$) that are combined to create the ensemble model $F_N(x)$.

\textbf{Query-only Transformer} This baseline, simply a VSI Transformer without any dense features, is added to evaluate the upper bound of a language only state-of-the-art classifier. 

\textbf{Late Fusion Transformer} This is an ablative baseline for the VSI Transformer. Instead of employing dense feature memory, we combine the dense features with the Transformer output at the final layers. Hence, we call this baseline \textit{Late Fusion}, representing how the fusion of modalities is done at the final stages.
Concretely, we concatenate the dense features to the pooled output from the transformer layer stack and then add a few layers of MLP before adding the classification head. See figure \ref{fig:architecture} for setup details.

\begin{table*}[h]
  \label{tab:f1_result}
  \small
  \begin{tabular}{llccc}
    \hline
    Model&Input Features&US & CA & GB\\
    \hline
    Query-only Transformer & Q only & 0.9395 / 0.9060 &0.8715 / 0.8059 &0.8896 / 0.8159\\
    AdaBoost 20 estimators &DF only & 0.8288 / 0.8399 &0.7830 / 0.7918 &0.8909 / 0.9019\\
    AdaBoost 50 estimators & DF only& 0.8570 / 0.8598 &0.8315 / 0.8678 &0.9132 / 0.9128 \\
    Transformer Late Fusion  & Q + DF & 0.9780 / 0.9698  & 0.9585 / 0.9289 & 0.9719 / 0.9519 \\
    \hline
    VSI Transformer & Q + DF & \textbf{0.9868} / \textbf{0.9809} & \textbf{0.9784} / \textbf{0.9655} & \textbf{0.9824} / \textbf{0.9730} \\
   \% Improvement (vs Query-only) & - & +5.0\% / 8.3\% & +12.3\% /  +19.8\% & +10.4\% / +19.3\% \\
   \% Improvement (vs Adaboost) & - & +15.1\% / +14.1\% & +17.7\% /  +11.3\% & +7.6\% / +6.6\% \\
   \% Improvement (vs best) & - & +0.8\% / +1.1\% & +2.1\% / +3.9\% & +1.1\% / +2.2\%  \\
  \hline
\end{tabular}
  \caption{F1 and Precision metrics on COVID-19 vaccination access search intent prediction. VSI Transformer outperforms best Transformer baseline by $+0.8\%$ to $+2.1\%$ and Adaboost by up to $+17.7\%$ F1 score.}
\end{table*}


\begin{table}[ht]
\small
  \begin{tabular}{lcccc}
    \hline
    Model & N$_{\ell}$ & US & CA & GB\\
    \hline
    Late Fusion & 1 & 0.9780 & 0.9585 & 0.9719 \\
    Late Fusion & 2 & 0.9827 & 0.9690 & 0.9817 \\
    Late Fusion & 3 & 0.9846 & 0.9788 & 0.9822 \\
    VSI & 1 & 0.9868 & 0.9784 & 0.9824 \\
    VSI & 2 & 0.9870 & 0.9785 & 0.9840 \\
    VSI & 3 & \textbf{0.9872} & \textbf{0.9789} & \textbf{0.9848} \\
  \hline
\end{tabular}
  \caption{Impact of increasing the number of layers ($N_\ell$) of the MLP networks on F1.}
  \label{tab:f1_result_nl}
\end{table}

\subsection{Results \& Analysis}
Table \ref{tab:f1_result} presents the F1 and precision metrics from the vaccine intent classification tasks. VSI Transformer outperforms Adaboost models by relative $+15.1\%$ gain on the US locale, $+17.7\%$ gain on the CA locale, and $+7.6\%$ gain on the GB locale on F1 metric. The gains against traditionally strong ML models are substantial and compelling. When compared with NLU-only approaches (e.g., query-only Transformer), VSI Transformer again strongly outperforms the baseline. Finally, there are modest (but consistently strong) gains against the best and strongest baseline considered of up to $+2.1\%$ F1 score.

\subsubsection{Importance of Text and Dense Features}
We study text-only NLU models and feature-only state-of-the-art ML based Adaboost models. Generally, it is not clear if NLU-only models outperform Adaboost models (or vice versa). Both modalities have their fair share of wins and loses across the three datasets and six metrics. To this end, we find that the well-established version of combining text and dense feature to outperform both Adaboost and the Query-only Transformer, signifying the importance of having both modalities for building a successful model.

\subsubsection{Dense Feature Memory vs Late Fusion} Late fusion is a well-established way to combine end-to-end deep learning with real world tabular features \citep{severyn2015learning,tay2017learning}.
Our results show that, while the Transformer with Late Fusion performs the best out of all baselines, the VSI Transformer still comfortably outperforms the Late fusion method 
, whereby we show that our proposed integration is more effective. In regards to the problem space and domain, this also seems to imply that a deeper fusion of text and dense features can be key in obtaining better model quality. 




\subsubsection{Increasing depth of MLP network}
\begin{figure*}
  \includegraphics[width=0.45\textwidth,height=8cm]{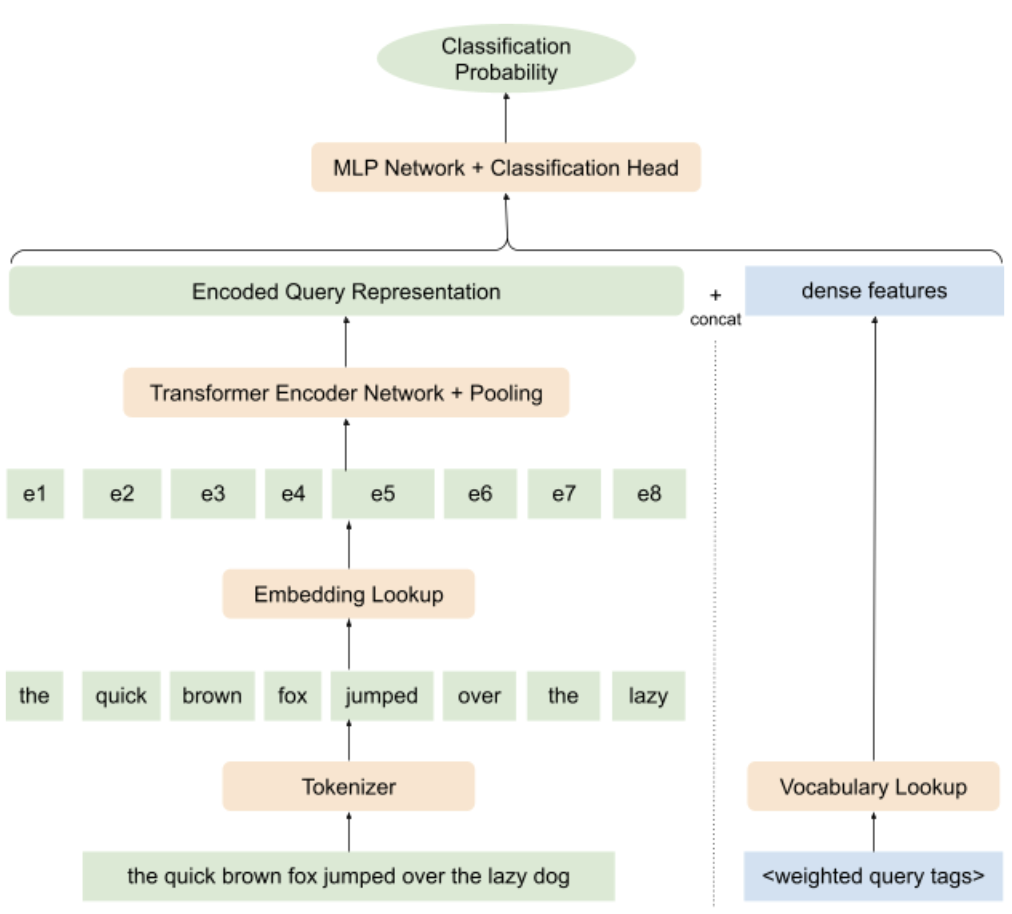}
  \hspace*{0.5in}
  \includegraphics[width=0.45\textwidth,height=8cm]{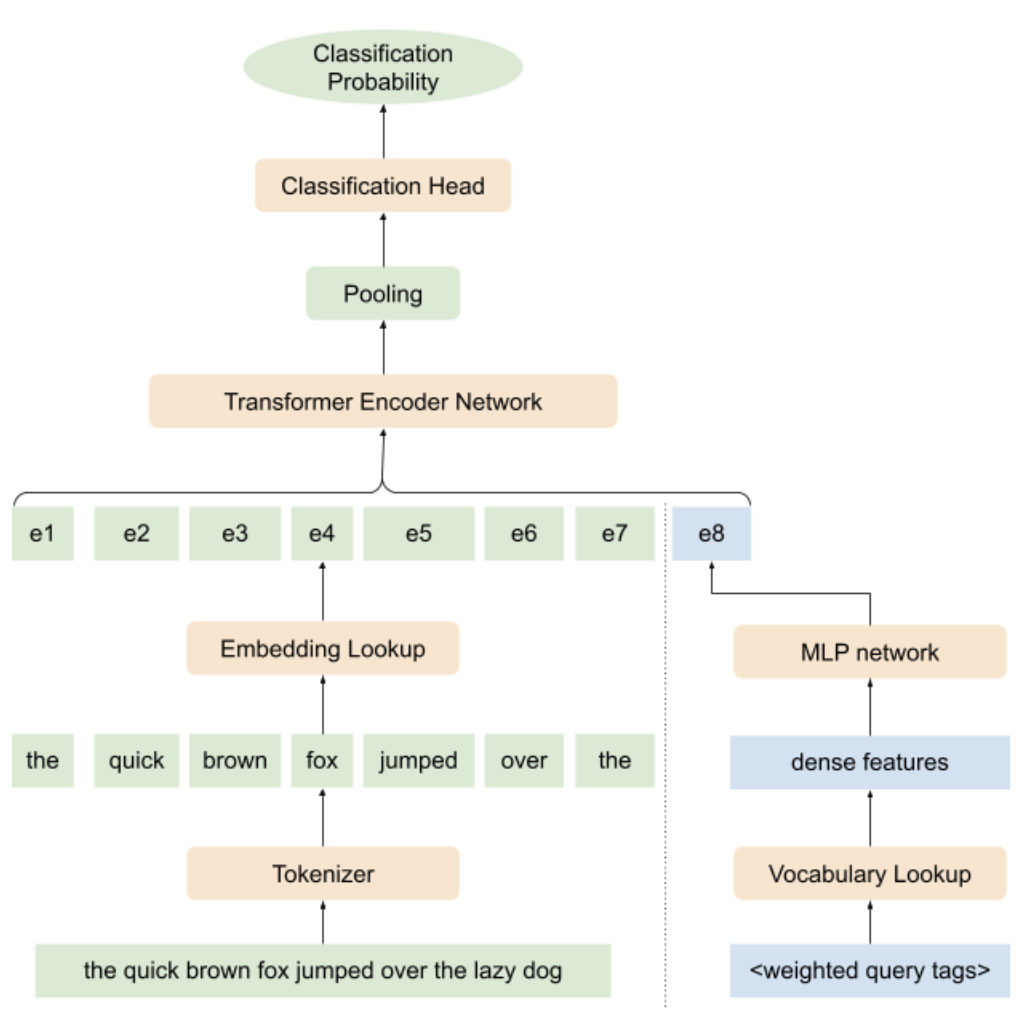}
  \caption{Depiction of adding transformed dense features to Transformer based models for a sample query. Architecture on the left depicts late fusion of the dense features to the query embeddings while the one on the right depicts addition of dense features to the query embeddings as a single memory token. Assumes that both the models have sequence length of 8 and the tokenizer produces one token for every word.}
    \label{fig:architecture}
\end{figure*}

The experiments described above used a single layer in the MLP networks in both the architectures present in Figure \ref{fig:architecture}. Adding a single layer means that we have almost the same number of additional parameters added in both the architectures for a fair comparison.

In this ablation (table \ref{tab:f1_result_nl}), we study the impact of using multiple layers in the MLP network. Note the each of these layers have a dimension size of 768 and use GeLU activation function. We see a consistent increase in performance as we increase the number of layers in the MLP network. This increase is more prominent in the Late Fusion architecture which starts to catch up (at the cost of increasing the depth of the model) but still performs worse than the VSI Transformer indicating that for the same number of model parameters, VSI Transformer is a better architecture.

\subsubsection{Using multiple memory tokens}

Table \ref{tab:mem_result} presents the results on increasing the number of memory tokens ($N_{memory}$) to up to 4 tokens. Though intuitively, it might seem that the performance will improve, experiments show that the correlation is not that straightforward.
\begin{table}[ht]
  \begin{tabular}{lccc}
    \hline
   $N_{memory}$ & US & CA & GB\\
    \hline
    0 & 0.9395 & 0.8715 & 0.8896\\ 
    1 & 0.9868 & 0.9784 & 0.9824 \\
    2 & 0.9875 & 0.9784 & 0.9850 \\
    3 & 0.9871 & 0.9788 & 0.9849 \\
    4 & 0.9853 & 0.9785 & 0.9839 \\
  \hline
\end{tabular}
  \caption{Trend of F1 metric on increasing the number of chosen memory tokens $N_{memory}$ in VSI transformer architecture.}
  \label{tab:mem_result}
\end{table}

The metrics seem to improve slightly but there is a consistent degradation observed with the F1 metric as we increase $N_{memory}$ to 4 tokens. When increasing $N_{memory}$ from 3 to 4 tokens, all three regions show a degradation.
Given the sequence length is constant, our hypothesis is that as we increase the number of memory tokens allocated to the dense features, we are using up tokens that could have otherwise been allocated to the query text. This reduces the length of query that can be seen by the model for large queries thereby impacting the overall performance of the model. Hence, in general, a single memory token might be sufficient, but the optimal number might differ from task to task.

\subsubsection{Analyzing improvement patterns}
Given the improvement in metrics, we looked at some of the examples where VSI Transformer model provides large gains over a query only model. An example pattern is "covid vaccine" followed by some noun. If the noun represents a location, the intent is to search for vaccine availability in that region. The query only Transformer model can only guess whether this represents a location unless it can recall from locations memoized during pretraining. Given that VSI transformer also uses dense features, it has additional signals that provide information like whether the query includes a location by sources like Google Cloud Entity analysis. Hence, using this additional signal, the model is able to cut down on a lot of false positives of this pattern.

\subsubsection{Using smaller Transformer models}
The above experiments are performed with T5 1.1 Base model, but models of this size are often prohibitive for online applications due to their resource requirements and latency where smaller and shallower models are preferred. But shallower models tend not to perform as good and one possible reason is that shallower models are limited in their capability to extract complex features.
The above experiments are performed with T5 1.1 Base model, but models of this size are often prohibitive for online applications due to their resource requirements and latency where smaller and shallower models are preferred. But shallower models tend not to perform as good and one possible reason is that shallower models are limited in their capability to extract complex features.

Given that dense features can provide complex features as inputs to the model in a preprocessed format, using dense features can provide a large boost in quality for shallow models. Additionally, such features can provide exclusive information that is not available in the text features (query). Hence, we performed an ablation study on usefulness of dense features on the model size.

Table \ref{tab:small_result} presents results on running the same experiments as above, but using T5 1.1 Small instead of the Base model. Overall, we see larger relative improvements compared to the Base model. For example, the VSI Transformer Small model has a relative F1 gain of 6.0\% and precision gain of 10.4\% over query-only Transformer Small model, much higher than the relative F1 gain of 5\% and precision gain of 8.3\% observed with the Base model.

\begin{table}[ht]
\small
  \begin{tabular}{lcc}
    \hline
    Model & Small & Base \\
    \hline
    Query-only & 0.9308 / 0.8866 & 0.9395 / 0.9060 \\
    \hline
    Late Fusion & 0.9539/0.9336 & 0.9780 / 0.9698 \\
    \% Imp. (vs Query-only) &+2.5\%/+5.3\% & +4.1\% / 7.0\% \\
    \hline
    VSI Transformer & 0.9862/0.9788 & 0.9868 / 0.9809 \\
    \% Imp. (vs Query-only) &+6.0\%/+10.4\% & +5.0\% / 8.3\% \\
  \hline
\end{tabular}
  \caption{Comparison of F1 and precision metric between T5 1.1 small and base models on US dataset.}
  \label{tab:small_result}
\end{table}

Even through shallow, the VSI Transformer's architecture is able to make better use of the attention layers for the dense features leading to pretty high boost even over the Late Fusion architecture, affirming that it is a better architecture for shallow models as well.

\section{Limitations}
Though we have shown that just one memory token is sufficient, assigning tokens to dense features means less number of tokens are available in the sequence for the text input. Another limitation is the use of locale specific vocabularies for dense features for each regional dataset as present in table \ref{tab:examples}, but that is not a limitation of the VSI Transformer but instead how the dense features are generated.

\section{Conclusion}
This paper presented an important task of classifying search queries with COVID-19 vaccination access intent. With an extensive set of experiments and comparing with strong baselines, we presented VSI Transformer, a novel and generic approach
that consistently and strongly outperforms all existing baselines that operate on either of the two modalities, or late fusion of the modalities. With an ablation study on model size, we show that for online applications where shallower models need to be deployed primarily due to latency constraints, making use of dense features can help bridge the gap in performance compared to deeper models. Future work in this direction can help further understand how to choose the optimal number of memory tokens, and explore more architectures to efficiently combine sequential and dense features.

\section*{Ethics Statement}
In our ongoing fight against the COVID-19 pandemic, understanding whether search queries exhibit an intent to seek vaccine access is an important problem to study to be able to collect search statistics about COVID-19 vaccination efforts. These statistics are made public via an online website, an interactive report that is accessible by anyone. All data used to train these models are anonymized and sampled, and labeled by a large pool of search quality raters who are trained to assess health-related information needs in search queries. These raters are unknown to and independent from the developers of the classifiers. This dataset is not made public or used in any other context.

\bibliography{anthology,custom}
\bibliographystyle{acl_natbib}

\appendix
\section{Appendix}
\label{sec:appendix}

\subsection{Implementation Details}
\label{sec:impldetails}
The VSI Transformer is implemented in Mesh Tensorflow \citep{shazeer2018mesh}, a Tensorflow-like API that supports distributed model parallelism. We initialize our model with the T5.1.1 Base checkpoint, comprised of $12$ encoder layers, $d_{model}$ size of 768, $d_{ff}$ of 2048. The model uses GEGLU-based feed-forward layers as described in \citep{shazeer2020glu}. The model has 12 heads. The overall number of non-embedding parameters is approximately $100M$ parameters. The model also utilizes the standard 32K SentencePiece that was trained on the C4 corpus. Our model is trained with 16 TPUv3 chips. We finetune all models using a sequence length of 32 subword tokens using the Adafactor optimizer \citep{shazeer2018adafactor}. The learning rate is a constant learning rate of $10^{-3}$ and the batch size is $128$. 

\subsection{Top Search Queries}
\label{sec:topq}
\begin{figure}[h!]
  \includegraphics[height=6cm]{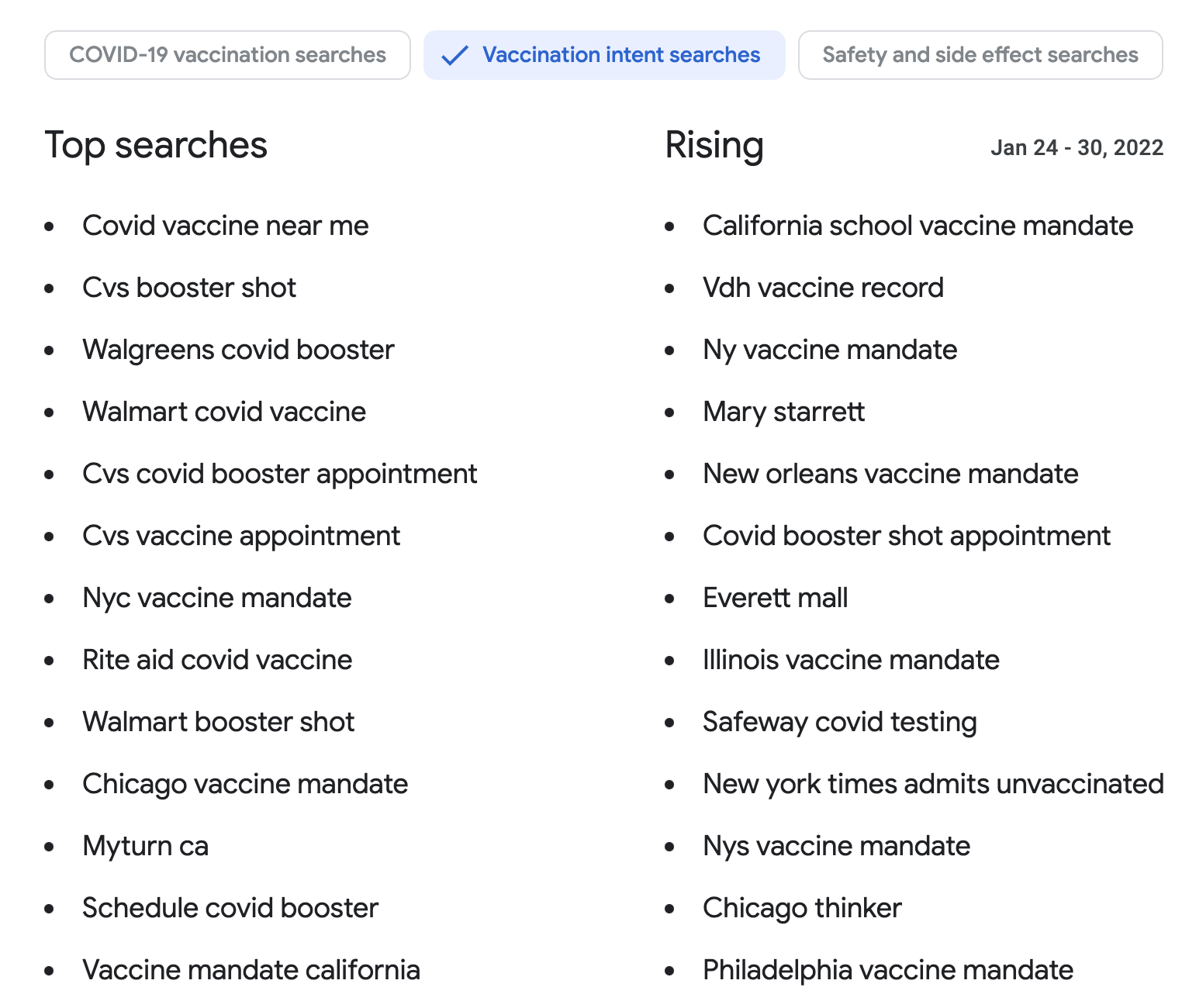}
  \caption{VSI tool presenting top search queries with vaccination intent}
    \label{fig:topq}
\end{figure}

Figure \ref{fig:topq} presents a view of the tool listing top search queries associated with vaccination intent.

\subsection{Sample Queries}
\label{sec:sample_queries}
Table \ref{tab:examples} presents a sample of some positive and negative queries for vaccine acces present in our dataset. The examples in the tabe shows how nuanced the problem is and why simply looking for termsatches didn't perform as good and a more sophisticated approach was needed.

\begin{table}[h!]
\centering
\small
\begin{tabular}{|p{1.6cm} |p{4.0cm} |}
 \hline
 \textbf{Class} & \textbf{Examples} \\ 
 \hline\hline
 Positive & 
 \begin{tabular}{l}
    \textit{covid vaccine appointment}\\
    \textit{where can i get covid vaccine} \\
    \textit{book covid jab} \\
    \textit{walk-in covid vaccine near me} \\
    \textit{nhs book covid vaccine} 
 \end{tabular} \\ 
 \hline
 Negative & 
 \begin{tabular}{l}
 \textit{covid stats by country}\\
 \textit{covid vaccine effectiveness} \\
 \textit{fully vaccinated travel}\\
 \textit{proof of covid vaccination}\\
 \textit{how long does the vaccine last}
 \end{tabular} \\
 \hline
\end{tabular}
\caption{Examples of positive and negative COVID-19 vaccine access queries.}
\label{tab:examples}
\end{table}

\subsection{Dense Feature Vocabulary}
As mention in section \ref{sec:dataset}, the dataset consists of dense features created from a vocabulary of 60,000 phrases. Table \ref{tab:top-features} presents a list of top phrases present in the vocabulary of each of the 3 regions.
\renewcommand{\arraystretch}{1.5}
\begin{table}[h!]
\centering
\small
\begin{tabular}{|p{1.6cm} |p{5.0cm} |}
 \hline
 \textbf{Category} & \textbf{Top Features} \\ 
 \hline\hline
 Vaccination access (US) & \textit{pharmacy, pfizer, vaccine appointment, appointment, pharmacies, moderna, dose, appointments, pfizer vaccine, cvs, walgreens, second dose, vaccine appointments, cvs pharmacy, doses, shot, cvs covid, walgreens pharmacy, vaccine eligibility, moderna vaccine} \\ 
 \hline
 Vaccination access (GB) & \textit{appointment, appointments, book, booking, vaccination centre, clinic, vaccination centres, vaccine appointment, clinics, nhs uk, nhs, coronavirus covid, walk in, coronavirus vaccination, covid vaccination, vaccination clinic, vaccine clinic, centres, vaccination appointment, vaccine centre, centre, book covid, pfizer, astrazeneca} \\
 \hline
 Vaccination access (CA) & \textit{clinic, appointment, clinics, vaccine clinic, vaccine appointment, vaccination clinic, pharmacy, appointments, book, booking, pharmacies, registration, pfizer, drug mart, vaccination center, walk in, vaccination appointment, vaccination clinics, vaccine pop up} \\
 \hline
\end{tabular}
\caption{Top features associated with a query in the dense feature vector, for each split.}
\label{tab:top-features}
\end{table}
\renewcommand{\arraystretch}{1.3}

\end{document}